\documentclass[conference]{IEEEtran}
\usepackage{epsfig,amsfonts,amsbsy,bm,mathrsfs}
\usepackage{amssymb,amsmath,amsthm,latexsym,amscd,amsfonts}
\usepackage{authblk}
\usepackage{graphics}
\usepackage{graphicx}
\usepackage{psfrag,float}
\usepackage{pstricks}
\usepackage{pst-plot}
\usepackage{cite}
\usepackage{balance}
\usepackage{epsfig}
\usepackage{epstopdf}
\usepackage{bbm}
\usepackage{dsfont}
\usepackage{setspace}
\usepackage{float}
\usepackage{relsize}
\usepackage{tikz}
\usepackage{comment}
\usetikzlibrary{%
	arrows,%
	shapes.misc,
	shapes.arrows,%
	chains,%
	matrix,%
	positioning,
	scopes,%
	decorations.pathmorphing,
	shapes.geometric,
	shadows,
	decorations.pathreplacing,
	patterns
}
\usetikzlibrary{shapes,arrows}
\usepackage{pgfplots}
\pgfplotsset{compat=newest}
\pgfplotsset{plot coordinates/math parser=false} 
\newlength\figureheight 
\newlength\figurewidth 
\usepackage{tikzscale}
\usepackage{epstopdf}
\usepackage{soul}
\usepackage{subfigure}

\newtheorem{remark}{Remark}

\begin{document}
	%
	%
	%
	%
	\title{3D Orientation Estimation with Multiple 5G mmWave Base Stations}
	\author{
		Mohammad A. Nazari\IEEEauthorrefmark{1}, Gonzalo Seco-Granados\IEEEauthorrefmark{2}, Pontus Johannisson\IEEEauthorrefmark{3}, Henk Wymeersch\IEEEauthorrefmark{1}\\
		 \IEEEauthorrefmark{1}Department of Electrical Engineering, Chalmers University of Technology, Sweden \\
		\IEEEauthorrefmark{2}Department of Telecommunications and Systems Engineering, Universitat Autonoma de Barcelona, Spain\\
		\IEEEauthorrefmark{3}RISE Research Institutes of Sweden, Sweden\\
		email: mohammad.nazari@chalmers.se
		}

	\maketitle
	
	%
	%
	%
	%
	\begin{abstract}
		We consider the problem of estimating the 3D orientation of a user, using the downlink mmWave signals received from multiple base stations. We show that the received signals from several base stations, having known positions, can be used to estimate the unknown orientation of the user. We formulate the estimation problem as a maximum likelihood estimation problem in the the manifold of rotation matrices. In order to provide an initial estimate to solve the non-linear non-convex optimization problem, we resort to a least squares estimation problem that exploits the underlying geometry. Our numerical results show that the problem of orientation estimation can be solved when the signals from at least two base stations are received. We also provide the orientation lower error bound, showing a narrow gap between the performance of the proposed estimators and the bound.
	\end{abstract}
	%
	%
	%
	%
	%
	%
	%
	%
	\section{Introduction}
	\label{sec:introduction}
	
	5G mmWave signals can provide accurate location information by virtue of their large bandwidth and large arrays at transmitter and receiver \cite{surveyRosado,2018-TCOM-ArshGboGnzHenk}. In particular, in contrast to 4G localization, which requires several synchronized base stations (BSs), in 5G mmWave, a single BS can be sufficient to obtain a location fix in 3D. 
	In order to localize using a single BS, related work on 5G mmWave localization assumed perfect synchronization between the user equipment (UE) and the BS \cite{2018-TCOM-ArshGboGnzHenk,2018-TWC-ZhrGnzHenk,2018-TWC-KakkavasNossek,2017-Glbcom-JukkaDestinoHenk,2017-arxiv-NLOS_components}. In practice, this assumption is not valid, prompting work on joint localization and synchronization by exploiting multipath information \cite{2018-Glbcom-HenkGnzSnw,mendrzik2019enabling,2020-sensor-YuGe}. In these works, the environment is mapped while at the same time the UE is localized and synchronized to the BS. 
	
	In addition to localization and synchronization, determining the orientation of the user is important for purposes of beamforming and beam tracking \cite{zhao2017angle,2019-ICC-BeamTracking}. Orientation information is obtained from angle measurements at the user side (i.e, angle of arrival (AoA) in downlink and angle of departure (AoD) in uplink). As was demonstrated in \cite{2018-TCOM-ArshGboGnzHenk}, with a single BS, the UE orientation in 2D can be determined when the UE is equipped with an array. In contrast, in 3D a single BS does not suffice to determine the UE orientation, since only angles in azimuth and elevation can be measured, unless additional signal sources are available, e.g., reflectors or scatterers \cite{2018-TWC-ZhrGnzHenk}, reconfigurable intelligent surfaces \cite{2020-arxiv-Alouini-RIS}, or additional BSs. Joint localization and orientation estimation was also considered in \cite{2018-ICC-Fundamental_Limits}, for anchor-free swarm navigation system, in \cite{2020-JSAC-LocalizationVLC} for visible light positioning, in \cite{2019-Asilomar-JointEstimation} for realistic channel realizations with hybrid array architectures, and in \cite{kakkavas2018multi} for  relative localization of vehicles. Orientation estimation from range measurements is also possible, as described in \cite{2014-TSP-Chepuri-Rigid_Body_Localization}. However, in the above contributions, the reliance on range measurements requires tight synchronization. To avoid this, pure angle-based localization or orientation estimation methods were pursued in \cite{wu2020cooperative}.

	 In this paper, we consider a mmWave MIMO scenario where the downlink signals are used to estimate the orientation of a UE with known position. While rotation estimation and tracking is a problem that has seen extensive treatment in the robotics literature \cite{barfoot2019state}, generally relying on an
	 inertial measurement unit (accelerometers and rate gyros), our formulation is unique as it provides \emph{absolute} 3D orientation information. 
	 The major contributions of this paper are as follows:
	    \begin{itemize}
	        \item We derive the orientation error bound for the estimation of the rotation matrix of a UE, using a constrained Fisher information analysis, which gives the lower bound for the performance of any unbiased estimator.
	        \item We pose a least squares (LS) optimization problem on the manifold of rotation matrices, giving a solution when the downlink signals from at least two BSs yield AoA measurements at the UE.
	        \item The obtained LS estimate can then be refined using a maximum likelihood optimization on the manifold of rotation matrices, leading to a solution with lower root mean squared error, and approaching the orientation error bound. 
    \end{itemize}
	    Our results show that the proposed estimation algorithms are efficient, approaching the orientation error bound, with low complexity.
	   
	\subsubsection*{Notations}
	We denote vectors and matrices with bold lowercase and uppercase letters ($\mathbf{x}$ and $\mathbf{X}$), respectively. The matrix $\mathbf{X}$ transpose is represented as $\mathbf{X}^\top$. We write the Kronecker product as $\otimes$, and the cardinality of a set $\mathcal{M}$ as $|\mathcal{M}|$. 
	
	%
	%
	%
	%

	\section{Problem Statement}
	\label{sec:ProblemStatement}
	\subsection{System Model}
	
	We consider a scenario where there are $M$ base stations (BSs) with known positions and known orientations, in a global coordinate system, to be used for the estimation of the orientation a user equipment (UE), in the environment. We denote by $\mathbf{p}_{m}=[p_{x,m},p_{y,m},p_{z,m}]^\top$, $m=1,\cdots,M$ the positions of the base stations. 
	We further assume that the position of the UE is known and equal to $\mathbf{p}=[p_x,p_y,p_z]^\top$, while the orientation $\mathbf{o}=[\alpha,\beta,\gamma]^\top$ is unknown and to be estimated\footnote{The angels $\alpha$, $\beta$, and $\gamma$ are called Euler, or Tait-Bryan angles.}. 
	The UE orientation determines a local frame of reference, conveniently described by a $3 \times 3$ rotation matrix in the special orthogonal group $\mathrm{SO}(3)$: $\mathbf{R} \in \mathrm{SO}(3)$ (i.e., an orthogonal matrix, satisfying $\mathbf{R}^\top \mathbf{R}=\mathbf{I}_3$ and $\det(\mathbf{R})=+1$). While angles $\alpha$, $\beta$, and $\gamma$ have physical meaning, they can be  related to the rotation matrix $\mathbf{R}$. 
	Accordingly, we mean \emph{estimating the rotation matrix $\mathbf{R}$}  when referring to \emph{3D orientation estimation}.
	
    The rotation order is important when mapping between the Euler angles and the rotation matrix. We consider the sequence of rotations around $z$, $y^\prime$, and $x^{\prime\prime}$, as the following:
\begin{align}
    & \mathbf{R} = 
    \mathbf{R}_z(\alpha)
    \mathbf{R}_y(\beta)
    \mathbf{R}_x(\gamma), 
\end{align}
where $\mathbf{R}_z(\alpha)$ denotes a rotation of $\alpha$ radians around the Z-axis
\begin{align}
    \mathbf{R}_z(\alpha) = \left[\begin{array}{ccc}
\cos(\alpha) & -\sin(\alpha) & 0\\
\sin(\alpha) & \cos(\alpha) & 0\\
0 & 0 & 1
\end{array}\right],
\end{align}
$\mathbf{R}_y(\beta)$ shows a rotation of $\beta$ radians around the Y-axis
\begin{align}
    \mathbf{R}_y(\beta) = \left[\begin{array}{ccc}
\cos(\beta) & 0 & \sin(\beta)\\
0 & 1 & 0\\
-\sin(\beta)& 0 & \cos(\beta)
\end{array}\right],
\end{align}
and $\mathbf{R}_x(\gamma)$ indicates a rotation of $\gamma$ radians around the X-axis
\begin{align}
    \mathbf{R}_x(\gamma) = \left[\begin{array}{ccc}
1 & 0 & 0\\
0 & \cos(\gamma) & -\sin(\gamma)\\
0 & \sin(\gamma) & \cos(\gamma)
\end{array}\right].
\end{align}
The system model is visualized in Fig.~\ref{fig:my_label}.
	
	\subsection{Signal Model}

\begin{figure}[t]
    \centering
    \includegraphics[width=1\columnwidth]{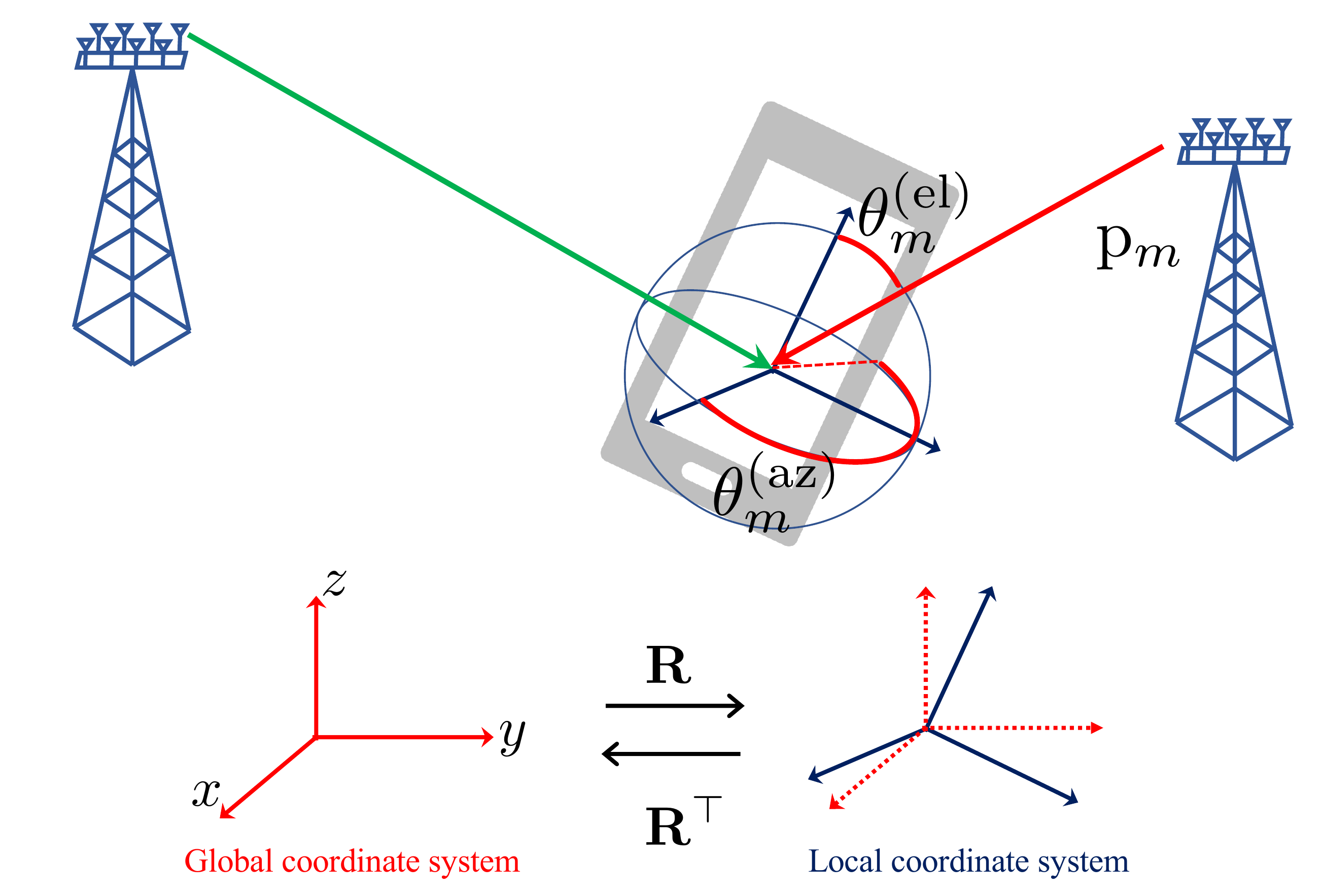}
    \caption{Schematic of 3D orientation estimation using downlink mmWave MIMO signals from 2 BSs.}
    \label{fig:my_label}
\end{figure}

	The BSs and the UE are equipped with arrays, which are capable of measuring the angles of departure (AoD) in the BS side, and angles of arrival (AoA) in the UE side.
	As a reference, we consider $\mathbf{R}=\mathbf{I}_3$ to correspond to the UE being parallel with the XY plane, with axis aligned with X and Y axis respectively.
	The signal observed by the UE is of the form
	\begin{align} \label{signal_model}
	    \mathbf{y}_{t}=\sum_{m=1}^{M} \alpha_m \mathbf{a}(\bm{\theta}_m)\mathbf{a}^\top(\bm{\psi}_m)\mathbf{s}_{m,t} + \mathbf{n}_{t},\,t=1,\ldots,T,
	\end{align}
	where $\alpha_m$ is the complex channel gain from BS $m$ to the UE, $\mathbf{a}(\bm{\theta}_m)$ is the UE response vector corresponding to AoA $\bm{\theta}_m=[\theta^{(\mathrm{el})}_m,\theta^{(\mathrm{az})}_m]^\top$ for elevation angle $\theta^{(\mathrm{el})}_m$ and azimuth angle $\theta^{(\mathrm{az})}_m$. Similarly, $\bm{\psi}_m$ denotes the AoD in elevation and azimuth from BS $m$.
	The transmitted signal by BS $m$ is  $\mathbf{s}_{m,t}$ (with power $P_m=\mathbb{E}\{\Vert\mathbf{s}_{m,t}\Vert^2\}$) and $\mathbf{n}_{t}$ is spatially and temporally white complex Gaussian noise with variance $N_0/2$ per real dimension. Since the UE location is known, the AoDs are known as well. 
	Under the considered model, it follows immediately that 
	\begin{subequations} \label{eq:AOA}
		\begin{align}
		    \theta^{(\mathrm{el})}_m & = \arccos\left( q_{z,m}/\Vert\mathbf{q}_m\Vert\right), \label{eq:defAOAel}\\
		    \theta^{(\mathrm{az})}_m & =\arctan 2(q_{y,m},q_{x,m}),\label{eq:defAOAaz}
        \end{align}
    \end{subequations}
with
\begin{align}\label{rotated_vectors}
    \mathbf{q}_m =  \mathbf{R}^\top (\mathbf{p}_m - \mathbf{p}).
\end{align}
The AoAs in the local coordinate system of the UE are also shown in Fig.~\ref{fig:my_label}.

	\subsection{Measurement Model}
	
	We assume an estimator exists that determines estimates of $\bm{\theta}_m$ from the observation $\mathbf{y}_t$, $t=1,\ldots, T$ (e.g., see \cite{2018-TCOM-ArshGboGnzHenk}). For simplicity of the exposition, we will model these AoA estimates as mutually independent with von Mises distributions, i.e., 
		\begin{align}
    p(\hat{\bm{\theta}}|\bm{\theta}) & = \prod_{m=1}^{M} \frac{1}{2\pi I_0(\kappa^{(\mathrm{el})}_m))} \exp(\kappa^{(\mathrm{el})}_m\cos(\hat{{\theta}}^{(\mathrm{el})}_m-{{\theta}}^{(\mathrm{el})}_m)), \label{eq:likelihooda}\\
    & \times \frac{1}{2\pi I_0(\kappa^{(\mathrm{az})}_m))}  \exp(\kappa^{(\mathrm{az})}_m\cos(\hat{{\theta}}^{(\mathrm{az})}_m-{{\theta}}^{(\mathrm{az})}_m)) \label{eq:likelihoodb}
    \\
    & \propto \exp(\bm{\kappa}^\top\cos(\hat{\bm{\theta}}-\bm{\theta}))
    \label{eq:likelihood}
\end{align}
	where $I_0(\cdot)$ is the modified Bessel function of order $0$, $\kappa^{(\mathrm{el})}_m$ is the concentration parameter of the $m$-th AoA in elevation, $\kappa^{(\mathrm{az})}_m$ is the concentration parameter of the $m$-th AoA in azimuth. In \eqref{eq:likelihood}, 	we have overloaded the notation for cosines and aggregated the AoAs, their estimates, and the corresponding concentrations in the $2M\times 1$ vectors $\bm{\theta}$, $\hat{\bm{\theta}}$ and $\bm{\kappa}$. The concentration parameters $\bm{\kappa}$ depend on the quality of the estimator. 
	It is important to note that the AoAs are obtained in the local frame of reference of the UE, which depends on the UE orientation $\mathbf{R}$. 
	
	%
	%
	%
	%
	
	\section{Fisher Information Analysis}
	\label{sec:FIM}

    \subsection{Background on (Constrained) Fisher Information}
	When estimating an unknown vector $\bm{\eta}\in \mathbb{R}^N$, constrained to lie on a manifold $\mathbf{h}(\bm{\eta})=\mathbf{0}$ defined by $K\ge 0$ non-redundant constraints, from an observation $\mathbf{y}$, the error covariance (under certain technical conditions) is lower bounded as \cite{1998-SPLetter-CRLB_constrained} 
		\begin{align} \label{cons_CRLB}
	    \mathbb{E}\left\{ (\bm{\eta}-\hat{\bm{\eta}})(\bm{\eta}-\hat{\bm{\eta}})^\top\right\} & \succeq  \mathcal{I}_{\mathrm{const}}^{-1}(\bm{\eta}),
	\end{align}
	where 
	\begin{align} \label{Constrined_OEB}
	    \mathcal{I}_{\mathrm{const}}^{-1}(\bm{\eta}) & = \mathbf{M} (\mathbf{M}^\top \mathcal{I}(\bm{\eta}) \mathbf{M})^{-1}\mathbf{M}^\top,
	\end{align}
	in which $\mathcal{I}(\bm{\eta})\in \mathbb{R}^{N\times N}$ is the  unconstrained Fisher information matrix  
	\begin{align}
	    [\mathcal{I}(\bm{\eta})]_{n,n'}=-\mathbb{E}\left\{ \frac{\partial^2}{\partial \eta_n \partial \eta_{n'}} \log p(\mathbf{y}|\bm{\eta})\right\},
	\end{align}
	and 
	 $\mathbf{M} \in \mathbb{R}^{N \times (N-K)}$ with $\mathbf{M}^\top \mathbf{M}=\mathbf{I}_{N-K}$,  satisfying 
	\begin{align}
	   \frac{\partial \mathbf{h}(\bm{\eta})}{\partial \bm{\eta}^\top} \mathbf{M} = \mathbf{0}_{K \times (N-K)},
	\end{align}
	is obtained by collecting the orthonormal basis vectors of null-space of the gradient matrix ${\partial \mathbf{h}(\bm{\eta})}/{\partial \bm{\eta}^\top} \in \mathbb{R}^{K\times N}$.
	Note that as a special case without constraints ($K=0$), we obtain the standard Fisher information matrix inequality $\mathbb{E}\left\{ (\bm{\eta}-\hat{\bm{\eta}})(\bm{\eta}-\hat{\bm{\eta}})^\top\right\} \succeq \mathcal{I}^{-1}(\bm{\eta})$.

	\subsection{Measurement FIM}
	Given a likelihood of the form \eqref{eq:likelihood}, the Fisher information of $\bm{\theta}$ is  
	\begin{align} \label{eq:FIMMeasurement}
	    \mathcal{I}(\bm{\theta})= \mathrm{diag}(\bm{\kappa} \odot I_1(\bm{\kappa}) \oslash I_0(\bm{\kappa}) ),
	\end{align}
	where $\odot$ and $\oslash$ denote pointwise product and division, respectively, and $I_1(\cdot)$ is the modified Bessel function of order $1$. The proof is provided  in Appendix \ref{app:FIMVonMises}.



	\subsection{Orientation FIM}
	
	To obtain the Fisher information of the rotation matrix $\mathbf{R}$, we 
	vectorize 
	$\mathbf{R}=[\mathbf{r}_1,\mathbf{r}_2,\mathbf{r}_3]$ as
	\begin{align}
	    \mathbf{r}=\mathrm{vec}(\mathbf{R})=[\mathbf{r}_1^\top,\mathbf{r}_2^\top,\mathbf{r}_3^\top]^\top.
	\end{align}
	The unconstrained Fisher information matrix $\mathcal{I}(\mathbf{r}) \in \mathbb{R}^{9 \times 9}$ of $\mathbf{r}$ is then obtained using the transformation matrix relating the measurements to the elements of the unknown rotation matrix $\mathbf{R}$ as in the following:
	\begin{align}
	    \mathcal{I}(\mathbf{r}) = \boldsymbol{\Upsilon} \mathcal{I}(\boldsymbol{\theta})\boldsymbol{\Upsilon}^\top,
	\end{align}
	where
	\begin{align}
	    \boldsymbol{\Upsilon} = \frac{\partial \bm{\theta}}{\partial \mathbf{r}}, 
	\end{align}
	with the obtained elements as in Appendix \ref{app:gradients}.
	However, to account for the orthogonality constraint of the rotation matrix, i.e., $\mathbf{R}^\top \mathbf{R}=\mathbf{I}_3$, we have $K=6$ constraints
	\begin{align}
	    \mathbf{h}(\mathbf{r}) = [& \Vert\mathbf{r}_1\Vert^2  - 1 ,  \mathbf{r}^\top_2 \mathbf{r}_1 , \mathbf{r}^\top_3 \mathbf{r}_1, \nonumber \\ & 
	    \Vert\mathbf{r}_2\Vert^2 - 1 , \mathbf{r}^\top_2 \mathbf{r}_3 , \Vert\mathbf{r}_3\Vert^2-1]^\top = \mathbf{0}_{6\times1}.
	\end{align}
	The following matrix $\mathbf{M}$ is an orthonormal basis for the null-space of the gradient matrix ${\partial \mathbf{h}(\bm{\eta})}/{\partial \bm{\eta}^\top}$ \cite{2014-TSP-Chepuri-Rigid_Body_Localization}:
	\begin{align} \label{matrix_M}
	    \mathbf{M} = 
	    \begin{bmatrix}
	    -\mathbf{r}_3 & \mathbf{0}_{3\times 1} & \mathbf{r}_2 \\
	    \mathbf{0}_{3\times 1} & -\mathbf{r}_3 & -\mathbf{r}_1 \\
	    \mathbf{r}_1 & \mathbf{r}_2 & \mathbf{0}_{3\times 1}
	    \end{bmatrix}.
	\end{align}6
	
	Finally, we can define the orientation error bound (OEB) as
	\begin{align} \label{OEB}
	    \mathrm{OEB} & =\sqrt{\mathrm{trace}(\mathcal{I}_{\mathrm{const}}^{-1}(\mathbf{r}))} \\
	    & \le  \sqrt{\mathbb{E}\left\{ \Vert\mathbf{r}-\hat{\mathbf{r}}\Vert\right\}} = \sqrt{\mathbb{E}\left\{\Vert \mathbf{R}-\hat{\mathbf{R}}\Vert^2_F\right\}},
	\end{align}
	where $\Vert.\Vert_F$ is the Frobenius norm.
	
	
	\section{Methodology}
	\label{sec:Methodology}

In this section, we first describe the general principle of optimization over the $\mathrm{SO}(3)$ manifold. Then, we describe a method to obtain an initial estimate of the UE rotation, based on a least squares criterion. This estimate is then refined through the maximum likelihood criterion.

\subsection{Optimization on the $\mathrm{SO}(3)$ Manifold}
To solve problems of the form 
\begin{align}
     \hat{\mathbf{R}} = \arg \min_{\mathbf{R} \in \mathrm{SO}(3)} f(\mathbf{R}),
\end{align}
where $f: \mathrm{SO}(3)\to \mathbb{R}$ is a smooth function, we rely on the method provided in  \cite[Chapter 4]{boumal2020introduction}. Starting from an initial estimate $\hat{\mathbf{R}}^{(0)}$, we compute 
\begin{align}
    \hat{\mathbf{R}}^{(k+1)}=\mathrm{Ret}_{ \hat{\mathbf{R}}^{(k)}}\left( -\varepsilon_k \mathrm{Proj}_{ \hat{\mathbf{R}}^{(k)}} \left.\frac{\partial f( \mathbf{R})}{\partial  \mathbf{R}} \right|_{\mathbf{R}=\hat{\mathbf{R}}^{(k)}} \right), 
\end{align}
where $\mathrm{Proj}_{\mathbf{X}}(\cdot)$ is a projection onto the tangent space (the set of real, skew-symmetric $3 \times 3$ matrices) at $\mathbf{X}$, $\mathrm{Ret}(\cdot)$ is a retraction from the tangent space onto $\mathrm{SO}(3)$, and $\varepsilon_k >0$ is a suitable step size. 
Intuitively, the gradient is calculated, projected to the tangent space (to follow the space of $\mathrm{SO}(3)$ as closely as possible), the initial matrix is updated, and then the updated matrix is normalized back into the $\mathrm{SO}(3)$ space. 
The projection and retractions operations are given by \cite[eqs.~(7.36) and (7.22)]{boumal2020introduction}
\begin{align}
    \mathrm{Proj}_{\mathbf{X}}(\mathbf{U})& =\mathbf{X}\mathrm{skew}(\mathbf{X}^\top \mathbf{U}),\\
    \mathrm{Ret}_{\mathbf{X}}(\mathbf{U}) & = (\mathbf{X}+\mathbf{U})(\mathbf{I}_3 + \mathbf{U}^\top \mathbf{U})^{-1/2},
\end{align}
where $\mathrm{skew}(\mathbf{Z})=(\mathbf{Z}-\mathbf{Z}^\top)/2$. It can be verified that $ \mathrm{Ret}_{\mathbf{X}}(\mathbf{U}) \in \mathrm{SO}(3)$ when $\mathbf{U}$ belongs to the tangent space at $\mathbf{X}\in \mathrm{SO}(3)$. 
Hence, optimization on the manifold requires definition of an initial estimate, the cost function $f(\mathbf{R})$ and its unconstrained gradient ${\partial f(\mathbf{R})}/{\partial \mathbf{R}}$.

	\subsection{Least Squares (LS) Estimation}

	According to \eqref{eq:AOA}, we can obtain an estimate $\hat{\mathbf{q}}_m(\hat{\bm{\theta}}_m)$ of ${\mathbf{q}}_m$ from the estimated AoA $\hat{\bm{\theta}}_m$ from  BS $m$ in azimuth and elevation, since all the distances to the different BSs are assumed known.
	Hence, when considering several such estimates, we can solve for $\mathbf{R}$ using the relation \eqref{rotated_vectors}.
	Hence, we use the following procedure. First, we select a subset $\mathcal{M} \subseteq \{1,\ldots,M\}$ of BSs. 
	  From $\hat{\bm{\theta}}_m$, $m\in\mathcal{M}$, we  compute $\hat{\mathbf{q}}_m(\hat{\bm{\theta}}_m)$ from the relations
	    \begin{align}
	         & \hat{q}_{z,m} = \Vert\mathbf{p}_m-\mathbf{p}\Vert \cos (\hat{\theta}^{(\mathrm{el})}_m),\\
	    & \hat{q}_{y,m} = \Vert\mathbf{p}_m-\mathbf{p}\Vert \sin (\hat{\theta}^{(\mathrm{el})}_m) \sin(\hat{\theta}^{(\mathrm{az})}_m),\\
	    & \hat{q}_{x,m} = \Vert\mathbf{p}_m-\mathbf{p}\Vert \sin (\hat{\theta}^{(\mathrm{el})}_m) \cos(\hat{\theta}^{(\mathrm{az})}_m).
	    \end{align}
 We then create a matrix $\mathbf{Q} \in \mathbb{R}^{3\times |\mathcal{M}|}$ that contains the estimates $\hat{\mathbf{q}}_m(\hat{\bm{\theta}}_m)$ as columns, so that $\mathbf{Q} \in \mathbb{R}^{3\times |\mathcal{M}|}$. Finally, we  solve the LS problem (initialized with the identity matrix)
	  	\begin{align} \label{LS_problem}
	    \hat{\mathbf{R}}_{\mathrm{LS}} = & \arg \min_{\mathbf{R} \in \mathrm{SO}(3)}  \Vert\mathbf{U}-\mathbf{R}\mathbf{Q}\Vert^2_F,
	\end{align}
	where $\mathbf{U}=\mathbf{P}-\mathbf{p}\otimes \mathbf{1}^\top_{|\mathcal{M}|}$, in which $\mathbf{P}=[\mathbf{p}_{m\in \mathcal{M}}]$. The gradient of the cost function is 
	\begin{align}
	    \frac{\partial \Vert\mathbf{U}-\mathbf{R}\mathbf{Q}\Vert^2_F}{\partial \mathbf{R}}  = -2 \Vert\mathbf{U}-\mathbf{R}\mathbf{Q}\Vert_F {\mathbf{Q}}^\top.
	\end{align}
	
	\begin{remark}
	The complexity of the method above grows with the number of used BS $\mathcal{M}$. To reduce the complexity, since $\mathbf{R}$ has only three degrees of freedom, at least 2 BSs should be used, so $|\mathcal{M}|=2$ suffices. These 2 BSs can be chosen based on their relative geometry (e.g., not colinear with the UE) and the concentration values in the von Mises likelihood.
	\end{remark}

	\subsection{Maximum Likelihood (ML) Estimation}
	The ML estimate of $\mathbf{R}$ is obtained by maximizing the log-likelihood function. 
	\begin{align}
	    \hat{\mathbf{R}}_{\mathrm{ML}} = \arg \min_{\mathbf{R} \in \mathrm{SO}(3)}  \underbrace{-\bm{\kappa}^\top\cos(\hat{\bm{\theta}}-\bm{\theta}(\mathbf{R}))}_{=f(\mathbf{R})},
	\end{align}
	initialized with the LS estimate \eqref{LS_problem}. 
 The gradient of the log-likelihood function is given by 
\begin{align}\label{derivative_vm}
    \frac{\partial f(\mathbf{R})}{\partial \mathbf{R}} = &-\sum_{m=1}^{M} {\kappa}^{(\mathrm{el})}_m  
    \sin(\hat{{\theta}}^{(\mathrm{el})}_m-{\theta}^{(\mathrm{el})}_m(\mathbf{R})) \frac{\partial {\theta}^{(\mathrm{el})}_m(\mathbf{R})}{\partial \mathbf{R}} \notag \\
     &-\sum_{m=1}^{M} {\kappa}^{(\mathrm{az})}_m  
    \sin(\hat{{\theta}}^{(\mathrm{az})}_m-{\theta}^{(\mathrm{az})}_m(\mathbf{R})) \frac{\partial {\theta}^{(\mathrm{az})}_m(\mathbf{R})}{\partial \mathbf{R}} .
\end{align}
The gradients $\frac{\partial}{\partial \mathbf{R}} {\theta}^{(\mathrm{el})}_i(\mathbf{R})$ and $\frac{\partial}{\partial \mathbf{R}} {\theta}^{(\mathrm{az})}_i(\mathbf{R})$ are provided in Appendix \ref{app:gradients}.

	%
	%
	%
	%

	\section{Numerical Results}
	\label{sec:results}
	In this section, we analyse performance of the proposed estimators, and compare it with the  OEB from \eqref{OEB}. We also show the orientation error bound for a range of UE orientations.
	\begin{figure}
	    \hfill
        \includegraphics[width=0.95\columnwidth]{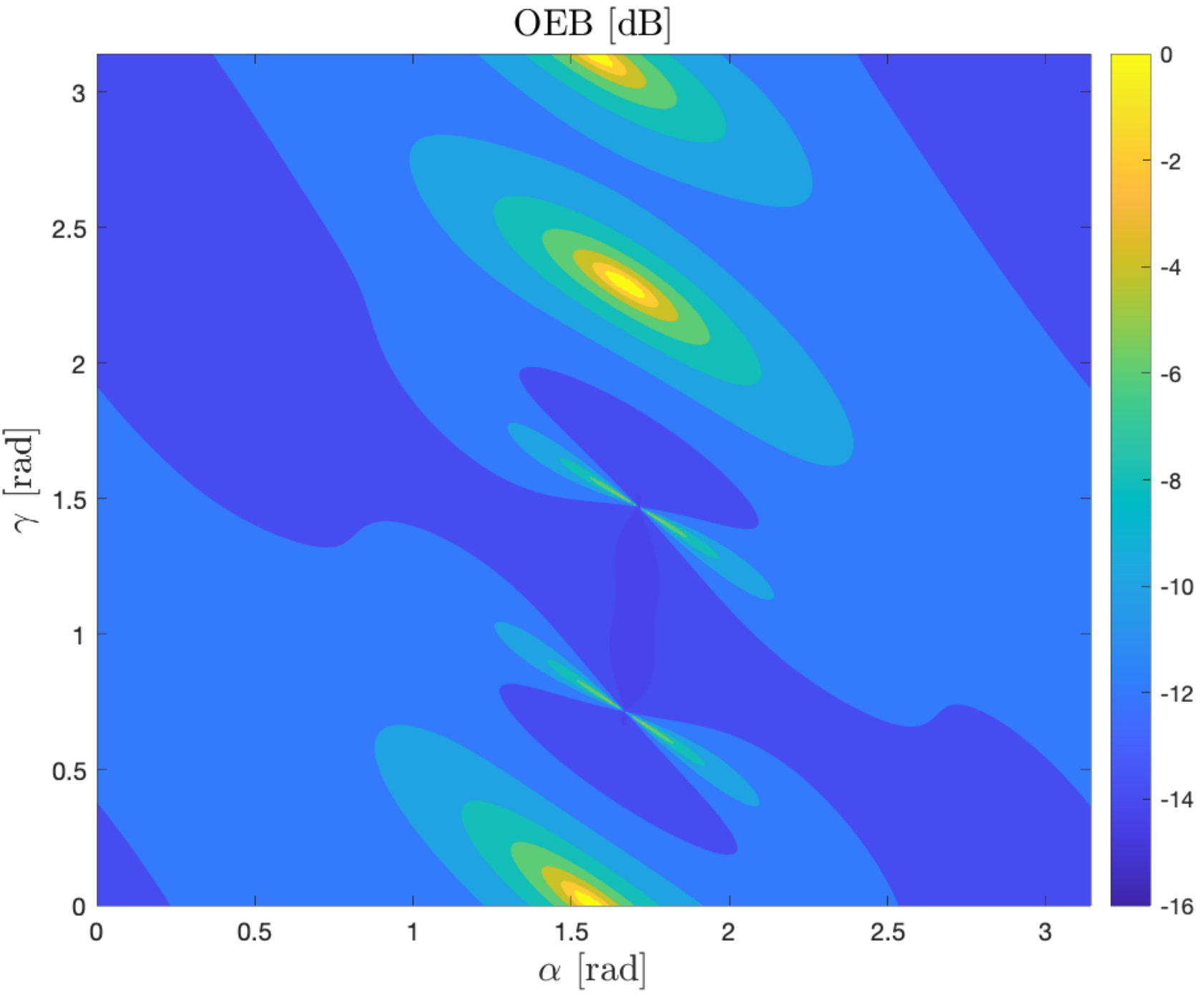}
        \caption{OEB in dB ($10 \log_{10}(\mathrm{OEB})$) for $\beta=-\pi/4$ vs. $\alpha,\gamma \in [0,\pi]$ with $M=2$ BSs.}
	    \label{fig:contour_plot_2BS}
	\end{figure}
    \begin{figure}
        \hfill
        \includegraphics[width=0.95\columnwidth]{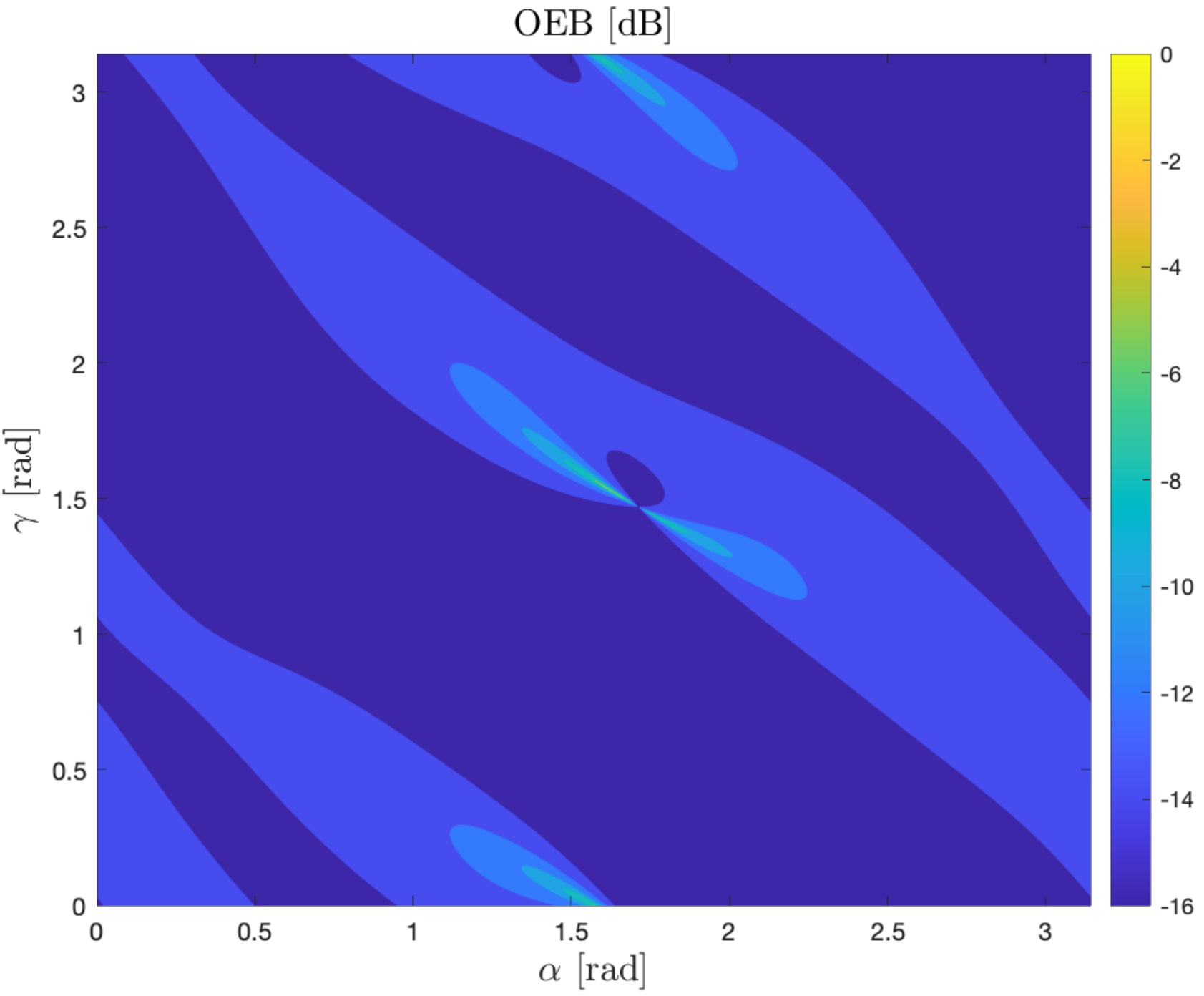}
        \caption{OEB in dB ($10 \log_{10}(\mathrm{OEB})$) for $\beta=-\pi/4$ vs. $\alpha,\gamma \in [0,\pi]$ with $M=3$ BSs.}
        \label{fig:contour_plot_3BS}
	\end{figure}

	\subsection{Simulation Scenario}
	Unless otherwise noted, we consider a scenario with $M=2$ BSs located at $\mathbf{p}_1=[0,0,0]^\top$ and $\mathbf{p}_2=[0,50,0]^\top$ sending downlink signals to the UE located at $\mathbf{p}=[50,0,-5]^\top$. The UE is equipped with a uniform planar array with $16 \times 16$ antenna elements with half-wavelength spacing. The carrier frequency is 28 GHz and the 
    transmitted signals are set as $\mathbf{s}_{m,t}=\sqrt{P_m}\mathbf{a}^{*}(\bm{\psi}_m)/\Vert \mathbf{a}(\bm{\psi}_m) \Vert $, so that we can define the SNR as
	\begin{align}
	    \mathrm{SNR}_m = \frac{|\alpha_m|^2 T N_{m,\mathrm{tx}} P_m}{N_0},
	\end{align}
    where $N_{m,\mathrm{tx}}$ denotes the number of transmit antennas at BS $m$, assuming coherent combining across transmissions. 
	Without loss of generality we set $\mathrm{SNR}_1=\mathrm{SNR}_2=\mathrm{SNR}$. The user orientation is set via the angles  $\mathbf{o}=[\alpha,\beta,\gamma]^\top$. 
	To obtain the measurements and their likelihoods, we proceed as follows:
	\begin{itemize}
	    \item We derive the FIM of $\mathcal{I}_{\bm{Y}}(\bm{\theta})$ from $\mathbf{y}_t$, $t=1,\ldots,T$ under the considered SNR. We use the subscript $\bm{Y}$ to express that the FIM is based on the received waveforms $\mathbf{y}_t$.
	    \item We equate $ \mathcal{I}^{-1}(\bm{\theta})$ to $\mathrm{diag}(\mathcal{I}^{-1}_{\bm{Y}}(\bm{\theta}))$.
	    \item We equate the appropriate diagonal elements in $ \mathcal{I}(\bm{\theta})$ to  ${\kappa}^{(\mathrm{el})}_m   I_1({\kappa}^{(\mathrm{el})}_m  )/I_0({\kappa}^{(\mathrm{el})}_m  )$ and ${\kappa}^{(\mathrm{az})}_m   I_1({\kappa}^{(\mathrm{az})}_m  )/I_0({\kappa}^{(\mathrm{az})}_m  )$, for $m =1,\ldots,M$, as derived in \eqref{eq:FIMMeasurement}. We then solve for ${\kappa}^{(\mathrm{el})}_m$ and ${\kappa}^{(\mathrm{az})}_m$. 
	\end{itemize}
	All manifold optimization problems were solved with the Manopt toolbox \cite{manopt}.
	
	\subsection{Results and Discussion}

	We first evaluate the impact of the orientation and the number of BSs on the OEB.  To visualize this, we fix $\beta$ to $-\pi/4$, and sweep $\alpha$ and $\gamma$ in the range $[0,\pi]$. The SNR is set to $-10~ \mathrm{dB}$. Fig.~\ref{fig:contour_plot_2BS} shows the corresponding result, with OEB values larger than 1 truncated. We observe low OEB for most UE orientations, but there are several peaks, where the OEB tends to infinity.  Specifically, when $\alpha \approx \pi/2$ and $\gamma \approx \pi/4$, the received ray from one of the BS hits the UE antenna array on the broadside, and does not provide a high quality orientation estimation. However, once this specific orientation changes, the downlink signal arrives at the UE array in a more suitable direction, facilitating a more satisfactory estimation, and accordingly lower OEB.
	{{Fig.~\ref{fig:contour_plot_3BS} depicts the OEB, when a third BS at the position $\mathbf{p}_3=[50,50,0]^\top$ is added. As observed, the OEB peaks are eliminated, and the error bound is greatly reduced for all considered orientations. The third BS can help to ensure that sufficient rays reach the UE with appropriate angles, not leading to unidentifiable orientation estimation.}}
    
	\begin{figure}
	    \centering
	    \footnotesize
	    \includegraphics{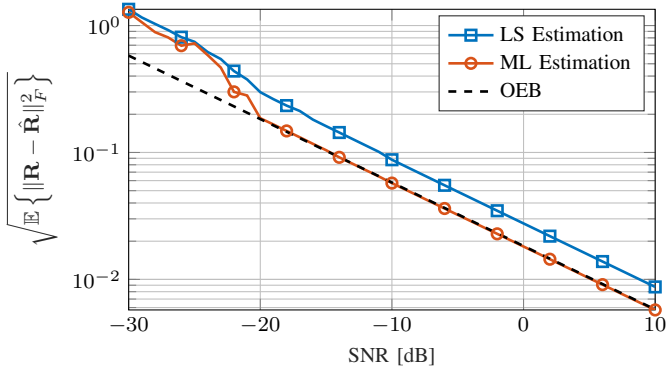}
	    \caption{Performance of Estimators and Comparison with the OEB vs. SNR.}
	    \label{fig:performance_plot}
	\end{figure}
	
	We now fix the orientation to  $[\alpha,\beta,\gamma]=[0.6\pi,0,-0.8\pi]$ and evaluate the performance of the proposed estimators as a function of the SNR. From 200 Monte Carlo simulations, we obtain an estimate of   $({\mathbb{E}\{\Vert \mathbf{R}-\hat{\mathbf{R}}\Vert^2_F\}})^{1/2}$, for both LS and ML estimators. We also plot the orientation error bound \eqref{Constrined_OEB}. 
		As observed, the performance of both estimators improves as SNR increases, which is expected. The gap between the performance of ML and LS is due to the fact that in LS, the estimation  neglects the distribution of measurements (i.e., the concentration of the von Mises distribution). The gap between the ML and LS estimator depends on the values of $\kappa$ for each AOA. When all $\kappa$ values are identical, the gap between ML and LS becomes smaller. 
		When it comes to comparison of the OEB with our proposed estimators, it is observed that the performance of ML estimator achieves the OEB for $\mathrm{SNR}>-20~\mathrm{dB}$.  This shows the efficiency of the proposed algorithms.

	%
	%
	%
	%
	\section{Conclusions}
	\label{sec:Conclusion}
	In this paper, we studied the problem of estimation of the orientation of a UE in 3D using downlink mmWave MIMO signals from multiple base stations.  
	The UE has an array of antennas which allows measuring the AoAs from the BSs, and solving for the unknown 3D orientation. Rotations are represented with rotation matrices in $\mathrm{SO}(3)$, which allows us to formulate and solve LS  and ML estimation problems on the manifold of 3D rotations. The solution of the LS problem was used as an initialization to the ML estimation problem, which is a non-convex optimization on the manifold $\mathrm{SO}(3)$. The performance of the resulting ML estimate coincides with the OEB, which is obtained by using the FIM of the rotation matrix subject to its orthogonality constraint.
	Future research would include the case where the position of the UE is unknown, and the study of dynamic situations with time-varying UE position and orientation. 
	
	\section*{Acknowledgment}
	This work was supported under the Wallenberg AI, Autonomous Systems and Software Program (WASP), the Swedish Research Council under grant 2018-03701, 
the Spanish Ministry of Science, Innovation and Universities under Projects TEC2017-89925-R and
PRX18/00638 and by the ICREA Academia Programme.
	
	%
	%
	%
	%
	\appendices
\section{FIM of Von Mises Distribution} \label{app:FIMVonMises}

We focus on one AoA, allowing us to remove all indices. The log-likelihood function is then given by
\begin{align}
    \log p(\hat{\theta}|\theta) &= \kappa \cos(\hat{\theta}-\theta) - \ln 2\pi I_0(\kappa).
\end{align}
Then we have 
\begin{align}
    \mathcal{I}(\theta) &= \mathbb{E}\left(-\frac{\partial^2}{\partial \theta^2}\log p(\hat{\theta}|\theta) \right) \\
    &= \mathbb{E}\left(\kappa \cos(\hat{\theta}-\theta) \right)\\
    &= \int_{\theta-\pi}^{\theta+\pi} \kappa \cos(\hat{\theta}-\theta) \frac{\mathrm{e}^{\kappa \cos(\hat{\theta}-\theta)}}{2\pi I_0(\kappa)} \mathrm{d}\hat{\theta}. \label{fimntegral}
\end{align}
By change of the variable $\hat{\theta}$, the equation \eqref{fimntegral} is simplified as 
\begin{align}
    \mathcal{I}(\theta) &= \int_{-\pi}^{\pi} \kappa \cos(\hat{\theta}) \frac{\mathrm{e}^{\kappa \cos(\hat{\theta})}}{2\pi I_0(\kappa)} \mathrm{d}\hat{\theta},
\end{align}
which in turn is analogous to the first-order Bessel function 
\begin{align}
    I_1(\kappa) = \int_{0}^{\pi} \frac{1}{\pi} \cos(\hat{\theta}) \mathrm{e}^{\kappa \cos(\hat{\theta})} \mathrm{d}\hat{\theta}.
\end{align}
Since $\cos(\hat{\theta})=\cos(-\hat{\theta})$, one can write
\begin{align}
    \mathcal{I}(\theta) 
    &= 2\int_{0}^{\pi} \kappa \cos(\hat{\theta}) \frac{\mathrm{e}^{\kappa \cos(\hat{\theta})}}{2\pi I_0(\kappa)} \mathrm{d}\hat{\theta} \\
    &= \frac{\kappa}{I_0(\kappa)} \int_{0}^{\pi} \frac{1}{\pi} \cos(\hat{\theta}) \mathrm{e}^{\kappa \cos(\hat{\theta})} \mathrm{d}\hat{\theta} \\
    &= \kappa \frac{I_1(\kappa)}{I_0(\kappa)}. \label{fim_von_mises}
\end{align}

\section{Gradient of the AoA with respect to the rotation matrix} \label{app:gradients}
	We introduce
\begin{align}
    \mathbf{u}^{(m)} &= \frac{\mathbf{p} - \mathbf{p}_m}{\Vert\mathbf{p} - \mathbf{p}_m\Vert},
\end{align}
and $\mathbf{u}_1 = [1,0,0]^\top$, $\mathbf{u}_2 = [0,1,0]^\top$,  $\mathbf{u}_3 = [0,0,1]^\top$.
This allows us to express \eqref{eq:defAOAel}--\eqref{eq:defAOAaz} as
\begin{align}
       \theta^{(\mathrm{el})}_m & = \arccos( -\mathbf{u}_3^\top \mathbf{R}^\top \mathbf{u}^{(m)}), \label{eq:defAOAel2}\\
		    \theta^{(\mathrm{az})}_m & =\arctan 2(-\mathbf{u}_2^\top \mathbf{R}^\top \mathbf{u}^{(m)},-\mathbf{u}_1^\top \mathbf{R}^\top \mathbf{u}^{(m)}),\label{eq:defAOAaz2}
\end{align}
We now make use of the following  identities 
\begin{align}
    \frac{\partial \mathbf{a}^\top \mathbf{X}^\top \mathbf{b}}{\partial \mathbf{X}} &= \mathbf{b} \mathbf{a}^\top,\\
    \frac{\partial}{\partial \mathbf{x}} \arccos ({u}(\mathbf{x}))&= -\frac{1}{\sqrt{1-{u}^2(\mathbf{x})}} \frac{\partial}{\partial \mathbf{x}} {u}(\mathbf{x}), \\
    \frac{\partial}{\partial \mathbf{x}} \arctan 2({u}(\mathbf{x}),{v}(\mathbf{x}))&= \frac{ {v}(\mathbf{x})\frac{\partial}{\partial \mathbf{x}} {u}(\mathbf{x})-{u}(\mathbf{x})\frac{\partial}{\partial \mathbf{x}} {v}(\mathbf{x})}{{u}^2(\mathbf{x})+{v}^2(\mathbf{x})},
\end{align}
in order to write 
\begin{align}
\frac{\partial \theta^{(\mathrm{el})}_m}{\partial \mathbf{R}}  &= 
    \frac{\mathbf{u}^{(m)} \mathbf{u}_3^\top}{\sqrt{1-(\mathbf{u}_3^\top \mathbf{R}^\top \mathbf{u}^{(m)}))^2}}, \\
\frac{\partial \theta^{(\mathrm{az})}_m}{\partial \mathbf{R}}  &= 
    \frac{(\mathbf{u}_1^\top \mathbf{R}^\top \mathbf{u}^{(m)})\mathbf{u}^{(m)}\mathbf{u}_2^\top - (\mathbf{u}_2^\top \mathbf{R}^\top \mathbf{u}^{(m)})\mathbf{u}^{(m)}\mathbf{u}_1^\top}{(\mathbf{u}_1^\top \mathbf{R}^\top \mathbf{u}^{(m)})^2 + (\mathbf{u}_2^\top \mathbf{R}^\top \mathbf{u}^{(m)})^2}.
\end{align}
From this, we immediately obtain ${\partial \theta^{(\mathrm{el})}_m}/{\partial \mathbf{r}}=\mathrm{vec}({\partial \theta^{(\mathrm{el})}_m}/{\partial \mathbf{R}})$ and ${\partial \theta^{(\mathrm{az})}_m}/{\partial \mathbf{r}}=\mathrm{vec}({\partial \theta^{(\mathrm{az})}_m}/{\partial \mathbf{R}})$. 

\balance
	\bibliographystyle{IEEEtran}
	\bibliography{IEEEabrv,references} 
	
\end{document}